\documentclass[aps,twocolumn,prl]{revtex4}
\usepackage{epsfig}
\begin{document}

\title{Topological Frustration in Graphene Nanoflakes:\\ Magnetic Order and Spin Logic Devices}
\author{
Wei L. Wang$^{1}$, Oleg V. Yazyev$^{2,3}$, Sheng Meng$^{1}$,
Efthimios Kaxiras$^{1}$
}
\affiliation{
$^1$ Department of Physics and School of Engineering and Applied Sciences,
Harvard University, Cambridge, MA 02138 \\
$^2$ Ecole Polytechnique F\'ed\'erale de Lausanne (EPFL),
Institute of Theoretical Physics, CH-1015 Lausanne, Switzerland\\
$^3$Institut Romand de Recherche Num\'erique en Physique
des Mat\'eriaux (IRRMA), CH-1015 Lausanne, Switzerland
}
\date{\today}

\begin{abstract}
Magnetic order in graphene-related structures can arise from size effects or from
topological frustration. We introduce a rigorous classification scheme for
the types of finite graphene structures (nano-flakes) which lead to large net spin
or to antiferromagnetic coupling between groups of electron spins.
Based on this scheme, we propose specific examples of structures that can serve as the fundamental
(NOR and NAND) logic gates for the design of high-density ultra-fast spintronic devices.  We demonstrate, using ab initio electronic structure calculations, that these gates can in principle operate at room temperature
with very low and correctable error rates.
\end{abstract}


\maketitle

Magnetism in solids is typically associated with the presence of transition metal elements with
$d$ electrons. Experimental evidence has recently shown that magnetism can also arise in solids
composed of elements with only $sp$ electrons, as, for example, in bulk proton irradiated
graphite~\cite{Esquinazi03}. Theoretical studies addressing the issue of magnetism in carbon-based materials have focused on point defects
~\cite{Duplock04,Yazyev07,Palacios08} and reduced dimensionality~\cite{Son06,Yazyev08a,Ezawa07,Fernandez-Rossier07,Wang08,Bhowmick08}. These works have shown that magnetism can arise in various situations, for example, the antiferromagnetic order across the edges of zigzag-edged
graphene nanoribbons~\cite{Son06} and the large net spin in zigzag-edged triangular graphene nanoflakes~\cite{Ezawa07,Fernandez-Rossier07,Wang08}. On the other hand, magnetism from $sp$ electrons is not new in organic chemistry: high-spin states have been observed in synthesized organic molecules,
for example, conjugated polyradicals~\cite{rajca01}and triangulene derivatives~\cite{Allinson95}, where all conjugated $\pi$
electrons can not be paired simultaneously. At present, it is not yet clear if the magnetic properties of irradiated bulk graphite and those of high-spin organic molecules have the same origin. Understanding of the magnetic order and its origin in various finite graphene structures is of fundamental importance as well as of practical interest for application in spintronics~\cite{Wolf01} where carbon based materials have recently demonstrated their potential~\cite{Hueso07}.

In this letter, we show that a distinct origin of magnetism in finite graphene structures
is topological frustration of $\pi$ bonds,
which is a generalization of the simple counting rule that
governs magnetic order in organic molecules.
We use the notion of this topological frustration to derive a rigorous classification
scheme for arbitrarily shaped graphene
nanoflakes, depending on whether only one or both sublattices
of the graphitic structure are frustrated.
From this classification scheme we identify which nanoscale structures can give rise
to a strong antiferromangetic (AF) coupling and propose a specific example of a structure which
can serve as  the fundamental (NOR and NAND) spin logic gate.  Finally, we employ
first-principles electronic structure calculations to show that this type of gate
can operate at room temperature, an important prerequisite for the design of realistic,
high-density ultra-fast spintronic devices based on graphene.

A graphene nanoflake (GNF) is an arbitrarily shaped finite graphene fragment
consisting of hexagonal rings and bounded by a single
(non-self-intersecting) topological circuit, where all in-plane dangling $\sigma$-bonds at the edge are assumed passivated.
We start the general classification of GNFs with the widely used $p_z$ band Hubbard model,
where magnetic correlations are described through
the hamiltonian
\begin{equation}
    {\mathcal H} = - t \sum_{\langle ij \rangle, \sigma} c_{i\sigma}^\dagger c_{j\sigma} + U \sum_{i} n_{i \uparrow} n_{i \downarrow}.
\end{equation}
In the first term, the tight-binding part, the operators $c_{i\sigma}$
($c_{i\sigma}^\dagger$) annihilate (create) an electron at site $i$
with spin $\sigma = \uparrow, \downarrow$ and $t$ is the hopping
integral between the nearest neighbor sites $i$ and $j$. The honeycomb lattice of graphene is
bipartite, that is, any pair of bonded nearest neighbor carbon atoms consists of
one atom from each of the two interpenetrating sublattices,
commonly denoted as $A$ and $B$ (Fig. 1).
The second term of the hamiltonian describes electron-electron interactions via the
on-site Coulomb repulsion $U$ with $n_{i\sigma}= c_{i\sigma}^\dagger
c_{i\sigma}$ the number operator.
This interaction may trigger an instability in the low-energy electronic
states and produce spin polarized states to minimize the total
energy. Zero-energy (non-bonding) states in a half-filled $\pi$
sub-band are particularly prone to become polarized for $U > 0$.

The occurrence of zero-energy eigenstates in the tight-binding
hamiltonian for a GNF can be accounted for by a theorem
on hexagonal graphs \cite{Fajtlowicz05}. The number
of such states, called ``nullity'' ($\eta$),
is determined by the topology of the GNF
according to the equation $\eta =\alpha-\beta$,
where $\alpha$ and $\beta$ are the maximum numbers of non-adjacent vertices
and edges respectively. The latter is also called the maximum matching of the GNF graph
and satisfies the relations $\beta=\theta=\nu$,
where $\theta$ and $\nu$ are the numbers of positive and negative eigenstates respectively.
The sum $\alpha+\beta$ equals to $N$, the total number of carbon atoms in the GNF.
When $\eta=0$, $\beta=N/2$ and all carbon atoms can be connected by a set of non-adjacent pair-wise bonds, which is referred to as ``perfect matching'' indicating a perfect pairing of all $p_z$ orbitals.
Otherwise, $\eta=\alpha-\beta=N-2\beta>0$, which is the number of sites that are
left out by the best possible matching. The inability to simultaneously pair all $p_z$ orbitals is entirely attributed to the topology of the GNF, therefore can be called topological frustration.

\begin{figure}
\includegraphics[width=0.45\textwidth]{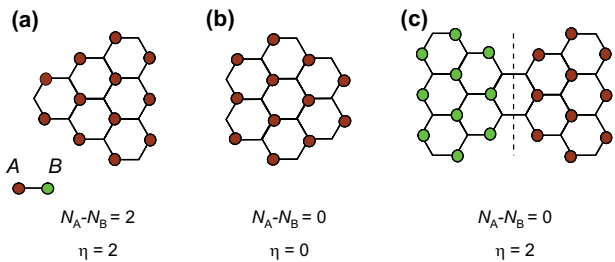}
\caption{\label{fig1} (Color online) (a) and (b) Class I GNFs, where nullity is equal to sublattice imbalance, $\eta = |N_{\rm A} - N_{\rm B}|$.
(c) Class II bowtie-shaped GNF with zero sublattice imbalance but a nullity of two, specifically,
$N_{\rm A}=N_{\rm B}=19$, $\alpha=20$.
The definition of $\alpha$ requires switching of sublattice across the dashed line. All colored sites correspond to a maximum set of non-adjacent sites.
}
\end{figure}

In general, GNFs
can be classified into two classes according to whether one or both
sublattices are topologically frustrated. In class I, at most one of the sublattices is frustrated,
which is characterized by $\beta=\min\{N_{\rm A}, N_{\rm B}\}$, that is,
the maximum matching covers all sites of at least one sublattice.
This class includes all highly symmetric forms of GNFs as we proved previously \cite{Wang08}.
The nullity for this class is simply $\eta=|N_{\rm A}-N_{\rm B}|$, as illustrated in Fig. \ref{fig1}(a)
and \ref{fig1}(b). Balanced sublattices ($N_{\rm A}=N_{\rm B}$) means zero nullity.
In class II, both sublattices are frustrated, characterized by
$\beta<\min\{N_{\rm A}, N_{\rm B}\}$ (note $\beta \leq \min\{N_{\rm A}, N_{\rm B}\}$),
which means $\eta>|N_{\rm A}-N_{\rm B}|$.
As a result, the nullity can be finite even for GNFs with balanced sublattices.
An example is shown in Fig.~\ref{fig1}(c),
where the nullity is $\eta=2$ even though $N_{\rm A}=N_{\rm B}$.

Although graph theory is all that is required to predict the number of singly-occupied orbitals, it
is not clear how the electron spins in these orbitals are aligned.
Complementary information comes from the Lieb theorem \cite{Lieb89}
which determines the total spin but not the number of
singly-occupied orbitals. The Lieb theorem was proved for any even-numbered bipartite
system, where the ground state has a total magnetic moment $S=|N_{\rm A}-N_{\rm B}|/2$.
Therefore, for class I GNFs, $S=|N_{\rm A}-N_{\rm B}|/2=\eta/2$, that is, all spins in singly-occupied orbitals align parallel to each other, consistent with Hund's rule. This was confirmed experimentally~\cite{Allinson95} and by first-principles calculations~\cite{Fernandez-Rossier07,Wang08}.
For class II GNFs, $S=|N_{\rm A}-N_{\rm B}|/2<\eta/2$,
indicating the existence of antiferromagnetic (AF) order, and Hund's rule breaks down. Specifically,
for the bowtie shaped GNF shown in Fig.~\ref{fig1}(c), the magnetic
moments of the left and right triangle must be AF coupled to
satisfy the requirement $S=0$. This is also consistent with the
fact that magnetic moments are localized in the two sublattices of
graphene favoring AF coupling~\cite{Yazyev07,Brey07}.

Besides topological frustration, AF coupling can also be induced by the polarization of the low-energy states that approach the Fermi level as the system size increases. This is a distinctly different origin of magnetism,
for two reasons: first, it can not give rise to net spin;
second, the energy of the non-interacting eigenstate is not strictly at the Fermi level,
except in infinite systems, and magnetic order appears only if the interaction
energy $U$ is above a positive threshold or, equivalently,
if the system is above a critical size.
Examples of this mechanism are
graphene nanoribbons \cite{Nakada96} and
hexagonal graphene nanoflakes \cite{Fernandez-Rossier07}.

We next focus on the magnetic coupling induced by topological
frustration in class II GNFs and use the bowtie structure of Fig.~\ref{fig1}(c) as the
simplest representatives. Such structures have a low-spin ground
state, involving spins spatially segregated and AF coupled. This
open-shell, low-spin feature is not only fundamentally
interesting but also may enable practically accessible logic
operations.  For instance, the simple bowtie structure
is a natural NOT gate because flipping the input spins on one side
of the bowtie requires the output spins on the other side to flip as well since
the spins on the two sides must point in opposite directions as long as the AF order is the ground state.
Practically, various means may be used to flip
the input spin, including polarized light, local magnetic fields or
direct injection of polarized electrons through magnetic materials.
Among those, spin injection~\cite{Wimmer08} appears the most promising, especially when
considering the natural integration of GNFs with
graphene nanoribbons which have been predicted to exhibit rich
spintronic properties including half metallicity~\cite{Son06,Wang08}.
The barrier for flipping the output spin is
expected to be extremely low ($\ll k_{\rm B} T$) due to the weak spin-orbit
coupling in carbon materials~\cite{Yazyev08a}, which is a
prerequisite for operation with low energy consumption.
In order to achieve both ultra-fast switching and robust operation at reasonable temperatures,
it is also important to have the magnetic coupling
$2J=E_{\rm FM}-E_{\rm AF}$, the energy difference between the ferromagnetic and
antiferromagnetic configurations, be greater than 18~meV,
the minimum energy dissipation~\cite{Landauer61} $kT_{\rm B}{\rm ln}2$ evaluated at 300~K.
In principle, picosecond flipping of electron spins can be achieved with an energy splitting
$J>\pi\hbar/10^{-12}$~sec~$\simeq 2$~meV.

\begin{figure}
\includegraphics[width=8cm]{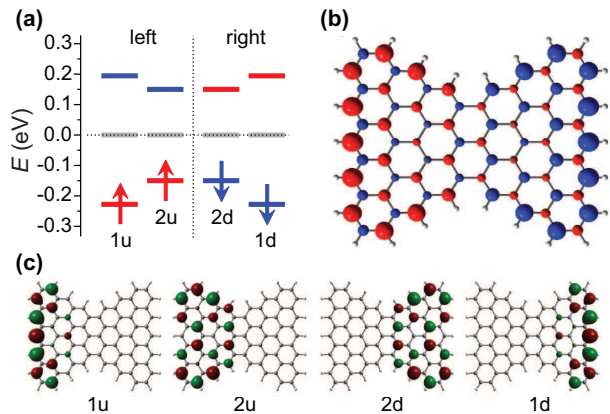}
\caption{\label{fig2} (Color online)
(a) The spectrum of singly occupied states of a bowtie-shaped GNF with
spin up (1u and 2u) and down (1d and 2d).
(b) Isodensity surface of the total spin distribution showing opposite spins
localized at opposite sides.
(c)~Wavefunctions of the four singly occupied states.}
\end{figure}

In order to establish whether or not the above conditions can be met,
we turn to first-principles electronic structure methods to
investigate the energetics of magnetic coupling in detail.
Calculations were performed
using spin-polarized density functional theory (DFT)
as implemented in the \texttt{SIESTA} code~\cite{SIESTA}. The generalized
gradient approximation (GGA) exchange-correlation functional~\cite{Perdew96}
was employed together with a double-$\zeta$ plus
polarization basis set, norm-conserving pseudopotentials
\cite{Troullier91} and a mesh cutoff of 200~Ry. The different spin
configurations were obtained by means of providing appropriate
initial guess electron spin densities~\cite{Yazyev08b,Convergence}.
The electronic structure of a representative bowtie GNF is shown in Fig. \ref{fig2}. The nullity of the corresponding
graph, $\eta=4$, manifests itself as four singly-occupied orbitals.
Within the spin-polarized formalism, when spin-spatial
symmetry breaking is allowed these four states are split, as shown in Fig.~\ref{fig2}(a).
The wave functions clearly illustrate the open shell singlet nature of the system: electrons
are AF coupled but not paired in the sense that they are spatially
segregated, in agreement with previous calculations~\cite{Pogodin03}
on non-kekul\'{e}an molecules. The spin coupling here is $2J=45$~meV,
well beyond the above mentioned thermodynamic threshold.

\begin{figure}
\includegraphics[width=0.45\textwidth]{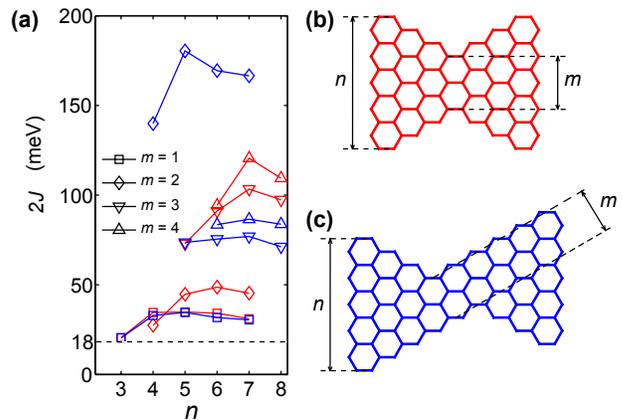}
\caption{\label{fig3} (Color online)
(a) Dependence of the spin-coupling $2J$ on the GNF bowtie geometry for (b) symmetric
and (c) asymmetric configurations of various triangular sizes $n$ and junction widths $m$.
The dashed horizontal line marks the minimum coupling
required for room temperature operation.
}
\end{figure}

The coupling strength in a bowtie GNF can be further engineered by optimizing its
geometry. To explore possibilities, we calculated
bowtie nanoflakes of different triangle size $n$ and junction width
$m$ (both measured in units of the
graphene lattice constant, $a_0=0.25$ nm) for both symmetric and asymmetric geometries, shown in Fig.~\ref{fig3}.
In either case, there are $n-m-1$ non-bonding states on each triangularly shaped side of the bowtie
GNFs. The calculated coupling magnitude $2J$
converges quickly with increasing $n$ after reaching a maximum
value Fig.~\ref{fig3}(a). The leveling off is attributed to the size effect
which diminishes the minimum energy splitting. Interestingly, for a specific asymmetric $m=2$ configuration, the interaction strengths are as large as $2J=180$~meV. In comparison, the AF coupling of quantum dots~\cite{Agarwal08} and transition
metal atoms~\cite{Hirjibehedin06} suffer from weak maximum coupling strength,
about 1 meV and 6 meV respectively, limiting their operation to very low temperatures.
With a coupling of 180 meV, a GNF-based spin gate may operate at room temperature
with an error rate of $p=e^{-2J/kT_{\rm B}}=0.001$, which can be handled by
error correction schemes.

\begin{figure}
\includegraphics[width=0.45\textwidth]{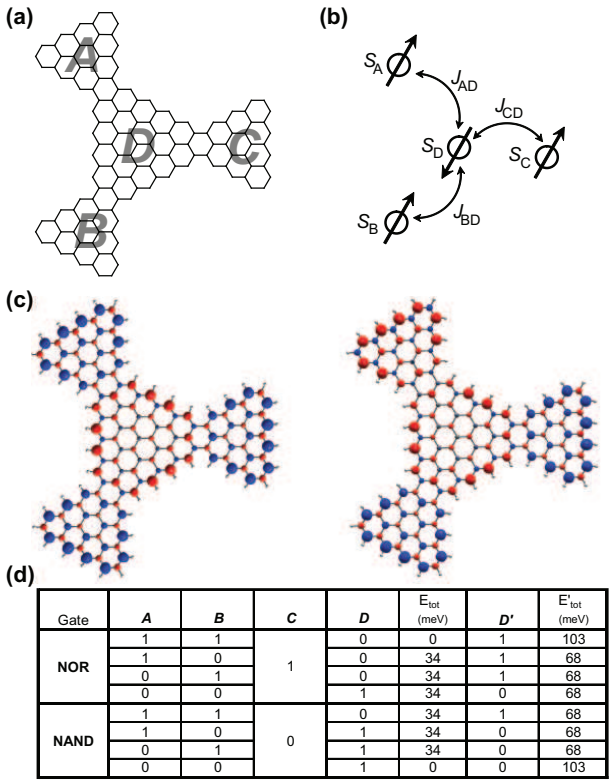}
\caption{\label{fig4} (Color online)
(a) Reconfigurable spin logic NOR and
NAND gate based on of a tri-bowtie GNF structure with $n_{\rm A}=n_{\rm B}=n_{\rm C}=4$,
$n_{\rm D}=6$, $m=1$ ($A$, $B$ and $D$ are two inputs
and one output, respectively, and $C$ is the programming bit).
(b) A scheme of the localized spins and the couplings ($2J_{\rm XY} =34$~meV).
(c) Two distinct spin configurations corresponding to 1110 and 0110 for
the $ABCD$ spins, respectively. (d)
The truth table of the programmable logic gate and the total energy
$E_{\rm tot}$ of the operation configuration;
$D'$ and $E'_{\rm tot}$ are the error output and the corresponding energy ($E'_{\rm tot}>E_{\rm tot}$).
}
\end{figure}

To further illustrate the concept of using Class II GNFs for spin logic processing,
we explore a tri-bowtie GNF in which the central triangle is connected
to three other surrounding triangles through
its vertices, as shown in Fig.~\ref{fig4}(a). From graph theory
arguments, we expect the number of unpaired spins in the central
region $D$ to be $n_{\rm D}-m_{\rm AD}-m_{\rm BD}-m_{\rm CD}-1$ where
$m_{\rm XY}$ ($X,Y=A,B,C,D$) are the widths of the junctions;
the peripheral triangles accommodate $n_{\rm A}-m_{\rm AD}-1$ unpaired spins.
The unpaired spins in the central region now
depend on the competition of the spins of all the three peripheral
regions: they tend to be AF coupled to the majority spins of the
peripheral regions so that the total energy is lowered. We can
therefore assign the total spins of two of those peripheral regions as
operands $A$ and $B$ and the third as a programming bit $C$; the spin in
the central region $D$ is the output. With the spin-up state representing
1 and the spin-down state 0, the above logic is written as
\begin{eqnarray}
D & = & \overline{(A \cap B)\cup(B \cap C)\cup(C \cap A)}
\nonumber \\
& = & \overline{(A \cap B)\cup((A \cup
B)\cap C)}
\nonumber
\end{eqnarray}
If $C$ is 1, the logic reduces to $D=\overline{A
\cup B}$, a NOR gate; if $C$ is 0, $D=\overline{A \cap
B}$, a NAND gate; these are the two fundamental gates in
Boolean logic, from either one of which all other gates can be
constructed. Therefore, the above scheme in principle provides an efficient
design, with the spin degree of freedom employed to satisfy in the
classic regime all digital logic operations.

Undoubtedly,  various engineering issues will have to be addressed
before the actual operation of such a device. For instance, the design of the
device ground state by coupling to peripheral leads, fan-out, and control of unidirectional logic flow are all open issues. Fabrication of the bowtie structure is another challenge, but recent experiments demonstrated that graphene devices only a few nanometers
in size can be sculpted by electron beam or scanning probes and are quite stable~\cite{Ponomarenko08,Tapaszto08}.
Meanwhile, we point out that the demands on fabrication are significantly alleviated
by the intrinsic defect-tolerability of the spin GNF devices:
at least $n-1$ out of $(n+2)^2-3$ carbon atoms can be removed (excluding those atoms at the junction to which the device function is highly sensitive) from the dominant sublattice in an individual triangle of size $n$ before the non-bonding states are eliminated and the local magnetic moment quenched.  These considerations indicate that
the devices based on the proposed design are not beyond the reach of modern
nano-scale fabrication methods.


\begin{thebibliography}{99}
\bibitem{Convergence}
In order to
obtain accurate total energies a 10$^{-5}$ maximum absolute change
of density matrix elements was used as a convergence criterion for
the self-consistent field (SCF) procedure. A Fermi-Dirac population
function ($k_{\rm B} T$=25~meV) common for both spin-up and spin-down
electrons was utilized. No constrains on the total spin of the
system were imposed during the SCF runs.
\end{thebibliography}
\end{document}